\documentclass[apj]{emulateapj}

\shorttitle{A SEARCH FOR YOUNG STARS IN S0 GALAXIES}
\shortauthors{JUST ET AL.}
\newcommand{\tx}[1]{\textrm{#1}}

\newcommand{\kms}{km~$\tx{s}^{-1}$}
\newcommand{\msun}{$\ensuremath{M_{\odot}}$}
\newcommand{\galex}{{\it GALEX}}

\def\gtsim{\lower 2pt \hbox{$\, \buildrel {\scriptstyle >}\over
    {\scriptstyle \sim}\,$}}
\def\ltsim{\lower 2pt \hbox{$\, \buildrel {\scriptstyle <}\over
    {\scriptstyle \sim}\,$}}
\journalinfo{The Astrophysical Journal, 2011}
\slugcomment{Accepted 2011 Jul 14}

\begin{document}
\title{A Search for Young Stars in the S0 Galaxies \\ of a Super-Group
  at $z=0.37$}
\author{Dennis~W. Just\altaffilmark{1},
  Dennis~Zaritsky\altaffilmark{1}, Kim-Vy~H. Tran\altaffilmark{2,3},
  Anthony~H. Gonzalez\altaffilmark{4},
  Stefan~J. Kautsch\altaffilmark{5,6}, John~Moustakas\altaffilmark{7}}
\altaffiltext{1}{Steward Observatory, University of Arizona, 933 North
  Cherry Avenue, Tucson, AZ 85721, USA}
\altaffiltext{2}{George P. and Cynthia W. Mitchell Institute for
  Fundamental Physics and Astronomy, Department of Physics, Texas A\&M
  University, College Station, TX 77843, USA}
\altaffiltext{3}{Institute for Theoretical Physics, University of
  Z\"{u}rich, CH-8057 Z\"{u}rich, Switzerland}
\altaffiltext{4}{Department of Astronomy, University of Florida, 211
  Bryant Space Science Center, Gainesville, FL 32611-2055, USA}
\altaffiltext{5}{Department of Physics, Computer Science \&
  Engineering, Christopher Newport University, 1 University Place,
  Newport News, VA 23606, USA}
\altaffiltext{6}{Division of Math, Science, and Technology, Nova
  Southeastern University, 3301 College Avenue, Fort Lauderdale, FL
  33314, USA}
\altaffiltext{7}{Center for Astrophysics and Space Sciences,
  University of California, San Diego, La Jolla, CA 92093, USA}


\begin{abstract}
  We analyze \galex\ UV data for a system of four
  gravitationally-bound groups at $z=0.37$, SG1120, which is destined
  to merge into a Coma-mass cluster by $z=0$, to study how galaxy
  properties may change during cluster assembly. Of the 38
  visually-classified S0 galaxies, with masses ranging from
  \hbox{$\log(M_*)[\msun]\approx10$--$11$}, we detect only one in the
  NUV channel, a strongly star-forming S0 that is the brightest UV
  source with a measured redshift placing it in SG1120. Stacking the
  undetected S0 galaxies (which generally lie on or near the optical
  red-sequence of SG1120) still results in no NUV/FUV detection
  ($<2\sigma$). Using our limit in the NUV band, we conclude that for
  a rapidly truncating star formation rate, star formation ceased {\it
    at least} $\sim0.1$ to $0.7$~Gyr ago, depending on the strength of
  the starburst prior to truncation. With an exponentially declining
  star-formation history over a range of time-scales, we rule out
  recent star-formation over a wide range of ages. We conclude that if
  S0 formation involves significant star formation, it occurred well
  before the groups were in this current pre-assembly phase. As such,
  it seems that S0 formation is even more likely to be predominantly
  occurring outside of the cluster environment.
\end{abstract}

\section{Introduction}\label{sec:intro}

S0 galaxies are more common in denser environments than in the field
\citep{Dressler80}, and the fraction of S0 galaxies increases over
time \citep{Dressler97,Fasano00,Desai07}, such that groups/clusters at
$z\sim0$ have S0 fractions $\approx3$ times greater than at
$z\sim0.5$. Due to the commensurate decline in the spiral fraction,
these findings have been interpreted as arising from the
transformation of spirals into S0's. Further observations have refined
the model to suggest that over this redshift range groups are the
primary site of S0 formation \citep[e.g.,][]{Wilman09,Just10},
i.e. the galaxies are ``preprocessed'' in groups prior to accretion
into the cluster \citep[e.g.,][]{Zabludoff96}.

However, the correlation between groups and clusters, and the
uncertainty in determining whether one is observing a group that will
soon fall into a cluster, complicate the interpretation of
environmentally dependent evolution. After all, galaxy properties
begin to change well outside of what is typically referred to as a
cluster \citep[i.e., 2 to 3 virial radii;][]{Lewis02,Gomez03}. The
question then becomes whether S0 formation occurs in isolated groups
or only when a group enters this meta-cluster environment. Is S0
formation related to the changes in star formation properties observed
in the far outskirts of clusters?

Super Group 1120-1202 (hereafter SG1120) provides a unique opportunity
to address these questions. It is a bound collection of four galaxy
groups at $z\sim0.37$ that is in the process of assembling into a
cluster. The four groups will merge by $z=0$ to form a cluster
one-third the mass of Coma or greater \citep{Gonzalez05}, yet they are
clearly independent groups as observed. Spectroscopic redshifts and
morphological classifications exist, allowing detailed analysis of its
constituent galaxies. The fraction of S0 galaxies in SG1120 is already
as high as that of clusters at similar redshift \citep{Kautsch08},
demonstrating that the high-density, massive cluster environment is
not the primary site of S0 formation. The question of whether these
S0's formed recently, in the pre-assembly epoch, is that which we now
consider.

To determine whether the S0's formed recently, we measure their recent
star formation history (SFH). A host of different mechanisms have been
suggested for the transformation, including mergers and galaxy-galaxy
interactions \citep{Toomre72,Icke85,Lavery88,Byrd90,Mihos04,Bekki11},
ram pressure stripping \citep{Gunn72,Abadi99,Quilis00}, strangulation
\citep{Larson80,Bekki02}, and harassment
\citep{Richstone76,Moore98}. The different mechanisms have their own
strengths and weaknesses. A difficulty with ram-pressure stripping as
the primary mechanism lies with accounting for the large fraction of
S0's in the field \citep[e.g.,][]{Dressler80,Dressler04}, although
ram-pressure stripping has been clearly observed in clusters
\citep[e.g.,][]{Irwin87,Kenney99} and could account for the deficiency
of HI gas observed in cluster spirals
\citep[e.g.,][]{vandenBergh76,Giovanelli83,vanGorkom96,vanGorkom04}. On
the other hand dynamical interactions (i.e. mergers and tidal effects)
are consistent with groups as the primary site of S0 formation
\citep[e.g.,][]{Wilman09,Just10}, although in this scenario it is
unclear why a comparable star-formation quenching efficiency is
observed in both groups and clusters \citep{Poggianti09b}. For an
excellent review of these different mechanisms and their ability to
explain observations across different environments and redshift, we
refer the reader to \cite{Boselli06}. These processes all involve the
halting of star formation, but operate on different timescales and
affect the SFH differently. By focusing on the SFH's of the S0's in
SG1120 we can constrain these mechanisms acting in a currently
assembling cluster.

Some measures of the SFH's of the S0's in SG1120 are already
available. Nearly all the S0's lie on or near the optical $B-V$ red
sequence (Figure~1) and inspection of their optical spectra reveal no
emission lines, suggesting no significant ongoing star
formation. Strong Balmer absorption indicative of star-formation
within the past $\sim1$~Gyr (so-called E$+$A galaxies; initial work by
\cite{Dressler83} and recent work, e.g., \cite{Yang08}) is also absent
in their spectra. However, all of these signatures are primarily
sensitive to significant bursts of recent star formation ($\sim$ tens
of percents by stellar mass). If the S0 formation process involves
more modest bursts (or just a truncation of a low level of star
formation), and if this happened recently ($<1$~Gyr ago), then
detection in the UV may be the best way to identify it. With these
goals in mind, we have obtained GALEX \citep{Martin05,Morrissey05}
imaging of SG1120.

Our paper is organized as follows. In \S2 we describe the data that
appear in this study. In \S3 we present our results, which we then
discuss and summarize in \S4. We adopt a cosmology with
$H_{0}=70$~\kms~Mpc$^{-1}$, $\Omega_{m}=0.3$, and
$\Omega_{\Lambda}=0.7$. Optical magnitudes are in the Vega system
while UV magnitudes are in the AB system; one can convert the $B$ and
$V$ magnitudes to the AB system by adding $-0.275$ and $-0.116$,
respectively.

\begin{figure}
\epsscale{1.0}
\plotone{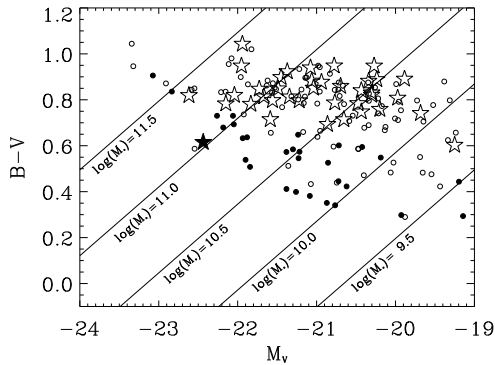}
\figurenum{1}
\caption{\label{optCMD} Rest-frame $B-V$ color magnitude diagram for
  spectroscopically confirmed SG1120 galaxies. S0 galaxies are
  highlighted as stars and the remaining members are shown as
  circles. NUV-detected galaxies are marked using filled-symbols, and
  approximate tracks of constant stellar mass are overplotted (see
  \S~\ref{sec:data}). Most S0's lie on the red sequence, consistent
  with being dominated by an old, passively evolving stellar
  population, and comprise $\sim35\%$ of all red sequence galaxies.}
\end{figure}


\section{Data}\label{sec:data}

Our analysis utilizes a combination of new and previously published
data, including GALEX, optical, and mid-infrared (MIR) imaging,
spectroscopy, and morphological classifications from high resolution
imaging.

In February 2009 we obtained {\it GALEX} imaging of SG1120 in both the
NUV and FUV bands\footnote{The NUV and FUV bands have effective
  wavelengths of $2271$\AA\ and $1528$\AA, respectively.}, with
exposure times of 31.5~ks and 33.0~ks, respectively. We generate
photometric catalogs using SEXtractor v2.8.6 \citep{Bertin96} with
matched apertures on the NUV and FUV images. We apply a detection
threshold on the NUV image of $2\sigma$ per pixel, with a minimum of 5
adjacent pixels required for a detection, and fix aperture radii at
5\arcsec\ (approximately twice {\it GALEX}'s FWHM). We identify UV
sources by cross-correlating the detections to galaxy optical
locations using a 1\arcsec\ matching threshold. We correct for
foreground galactic extinction with the \cite{Schlegel98} dust maps
and the \cite{ODonnell94} Milky Way extinction curve.

We utilize $B$ and $V$ band VLT/VIMOS photometry \citep{LeFevre03}
from \citet[hereafter T09]{Tran09}. Galactic extinction is corrected
for similarly as above \citep{ODonnell94,Schlegel98}. For the optical
data, we quote MAG\_AUTO magnitudes from SExtractor, which are similar
to Kron magnitudes (Kron 1980). While ideally one would want to use
PSF- and aperture-matched magnitudes when computing colors, we only
use the $B-V$ color for an estimate of stellar mass and to determine
whether a galaxy is blue or red. Stellar masses are determined
following the prescription of \cite{Bell03}, with the mass-to-light
ratios $(M_*/L)_B$ estimated using
\begin{equation}
  (M_*/L)_B=1.737(B-V)-0.942,
\end{equation}
assuming the diet Salpeter IMF defined in \cite{Bell01} and rest-frame
Vega magnitudes. Using a blue absolute magnitude of $M_B=5.45$ for the
Sun, a galaxy with $M_B=-19.5$ and $(B-V)=1$ has a stellar mass of
$\log(M_*)[\msun]=10.8$. Tracks of constant stellar mass are
overplotted in the color-magnitude diagram of Figure~\ref{optCMD}.

{\it Spitzer} imaging from MIPS \citep{Rieke04} that appears in T09 is
used for estimating MIR SFR's. T09 calculated ${\rm SFR}_{\rm IR}$ by
determining the total IR luminosity ($8$--$1000\mu m$) from the $24\mu
m$ luminosity using a family of IR spectral energy distributions
(SEDs) from \cite{Dale02}. Then, focusing on the SEDs representative
of the {\it Spitzer} Infrared Nearby Galaxies Survey \citep{Dale07}, a
median conversion factor was chosen at $z\sim0.37$ where the SEDs give
essentially the same values and the error is limited to
$\sim10$--$20\%$.

Spectroscopy for SG1120 come from VLT/VIMOS \citep[in
2003;][]{LeFevre03}, Magellan/LDSS3 (in 2006), and VLT/FORS2 \citep[in
2007;][]{Appenzeller98}, with resolutions of 2.5~\AA~pix$^{-1}$,
0.7~\AA~pix$^{-1}$, and 1.65~\AA~pix$^{-1}$, respectively. Further
details of the spectroscopic reduction can be found in \cite{Tran05}

\begin{figure}
\epsscale{0.8}
\plotone{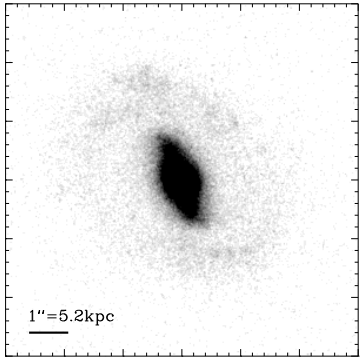}
\figurenum{2}
\caption{\label{S0img} {\it HST}/ACS F814W image of the UV-detected S0
  galaxy in SG1120, at \hbox{${\rm RA}=11^{\rm h}20^{\rm m}10.4^{\rm
      s}$}, \hbox{${\rm
      Dec}=-12^{\circ}01\arcmin51.7\arcsec$}. Classified as an S0,
  some structure is evident in the disk. The SFR derived from the MIPS
  and NUV images for this galaxy are 60~\msun~yr$^{-1}$ and
  20~\msun~yr$^{-1}$, respectively.}
\end{figure}

Morphological classifications exist for 143 of the
spectroscopically-confirmed SG1120 galaxies (T09) based on images
obtained with the {\it Hubble Space Telescope}/Advanced Camera for
Surveys ({\it HST}/ACS) in F814W ($11\arcmin \times
18\arcmin;0.05\arcsec$/pixel). Although with high-resolution HST
imaging it is possible to distinguish between elliptical and S0
galaxies \citep{Postman05}, some level of uncertainty in the
classifications exists regarding orientation angle \citep{Rix90},
surface brightness dimming, and the ``morphological $k$-correction''
\citep{Windhorst02,Papovich03}. The latter two effects tend to present
more difficulty for classifications of galaxies over a broad range of
redshifts, which is not the case for this study. The classification
scheme used by T09 assigns galaxies the average T-type visually
determined independently by four of the authors. We define our classes
as elliptical ($T \leq -3.5$), S0 ($-3.5 < T \leq 0$), and
spiral$+$irregular ($0 > T$). Thus, our definition of S0 spans S0/E to
S0/a. We require that at least one author classify a galaxy as S0
before it is included in our S0 sample; a combination of elliptical
and spiral classifications that average out to numerically meet our S0
criterion will not qualify as an S0. This definition results in 38
SG1120 galaxies classified as S0. This classification scheme is
different than that of \cite{Desai07} for the EDisCS sample
\citep{White05}, which can be used as a comparison sample, although
the primary difference is that for a given galaxy they assigned the
$T$-type most frequently assigned by their classifiers while we use
the average $T$-type. Adopting their classification scheme does not
change the results presented below.


\section{Results}\label{sec:resiults}
\subsection{UV Analysis}\label{sec:Si0results}

Of the 38 galaxies classified as S0, we detect one in the UV; it is
the brightest UV source among the spectroscopically identified
galaxies in SG1120, with $m_{NUV}=20.7$ and $m_{FUV}=21.5$
($>10\sigma$ detection in each band). This galaxy lies off the optical
red sequence as well and is detected at 24~$\mu$m (T09). Based on its
MIR and UV detections it has a significant amount of star-formation,
${\rm SFR}_{\rm IR}=60\pm12$~\msun~yr$^{-1}$ and ${\rm SFR}_{\rm
  UV}=21\pm2$~\msun~yr$^{-1}$, the latter of which is calculated from
its NUV magnitude without an intrinsic extinction correction using the
star-formation law of \cite{Kennicutt98}. There is structure apparent
in the disk of the galaxy (an {\it HST}/ACS F814W image of this galaxy
appears in Figure~\ref{S0img}), and given its strong SFR it is
possible that this is a misclassified spiral. While it could be
possible that we are missing a substantial population of blue S0's by
classifying such galaxies with disk structure as spirals, given the
already high S0 fraction in SG1120 it is unlikely that this is the
case.

\begin{figure}
\epsscale{1.0}
\plotone{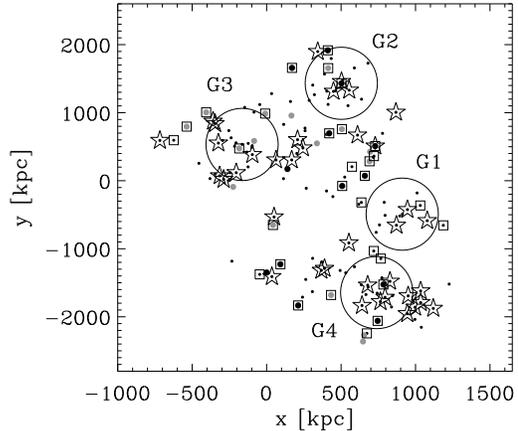}
\figurenum{3}
\caption{\label{spatplot} Spatial plot of the 143 SG1120 galaxies with
  morphological classifcations (dots). S0 galaxies are marked as
  stars, while galaxies detected in the NUV (gray circles) and
  NUV$+$FUV (black circles) are also highlighted. Galaxies from T09
  with ${\rm SFR}_{\rm IR}\ge 3$~\msun~yr$^{-1}$ based on MIPS data
  are marked with boxes.}
\end{figure}

The remaining 37 S0 galaxies are not detected in either the NUV or
FUV. Converting our UV detection limits to a SFR limit is not as
straightforward as above, since \cite{Kennicutt98} assumes a flat
spectrum from $1500$--$2800$\AA\ due to continuous star-formation for
longer than $100$~Myr, which need not be the case when we only have
upper limits on the UV emission. Therefore, we estimate the SFR upper
limit from the rest-frame $1500$\AA\ flux (which is less contaminated
from evolved stars than at $2800$\AA) after fitting the UV and optical
photometry of the S0's with {\tt KCORRECT} \citep{Blanton07}. This
results in a SFR limit of ${\rm SFR}_{\rm UV}\ltsim
0.1$~\msun~yr$^{-1}$ for the individual galaxies. While dust could
suppress NUV emission from star-formation, none of these S0's are
detected at 24~$\mu$m, although the MIR limit is weaker
($<3$~\msun~yr$^{-1}$; T09). A spatial plot showing the location of
the S0's, as well as the UV and MIR detections, appears in
Figure~\ref{spatplot}.

To look deeper for signs of recent or ongoing star-formation, we stack
the non-UV detected S0 galaxies. One of the S0's lies near the core of
Group 2, within 4\arcsec\ of a bright UV source (a star-forming
elliptical also detected at 24~$\mu$m with a SFR of
4.35~\msun~yr$^{-1}$; T09). Given the size of the GALEX PSF ($\approx
5\arcsec$), we exclude this source from the stacking analysis,
although its inclusion does not affect our results. We median stack
300 by 300 pixel thumbnails and compare the flux at the central
location to the distribution of fluxes in $\approx1500$
non-overlapping $5\arcsec$-radius apertures arranged such that they do
not touch the edge of the stacked image or the galaxy position. The
flux of the stacked S0 is $<2\sigma$ above the random fluctuations in
both the NUV and FUV, which corresponds to $m_{NUV}<26.0$ and
$m_{FUV}<26.6$ magnitude, and a SFR of $\ltsim0.01$~\msun~yr$^{-1}$
(estimated using the above method), an order of magnitude lower than
the constraint placed from the individual non-detections alone (see
above).\footnote{To derive a complementary SFR limit, we perform a
  similar stacking analysis with the MIPS data. However, given the
  crowded MIPS field $\sim$half of the S0 positions are contaminated
  with emission from nearby sources, making the interpretation of this
  result more difficult. In any event the limit inferred from this
  stacking is more than an order of magnitude weaker than the limit
  derived from the UV stacking.}

Early-type galaxies are known to have some UV emission,
i.e. ``UV-upturn'' galaxies \citep[e.g.,][and references
therein]{Greggio99,OConnell99,Brown03,Yi04}, which comes from evolved
stars. We compare our NUV detection limit with the model of
\cite{Han07}, who treat their model galaxy as a simple stellar
population with $\log(M_*)=10$. The expected NUV flux from evolved
stars for a $\log(M_*)=10.5$ galaxy (typical of the S0's in our
sample) is $\sim2$ magnitudes fainter than our stacked detection
limit.

\begin{figure}
\epsscale{1.0}
\plotone{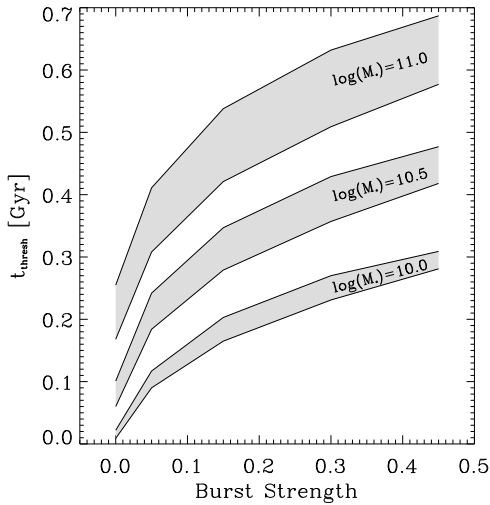}
\figurenum{4}
\caption{\label{scenarios} The time for a model galaxy to drop below
  our NUV detection threshold ($t_{\rm thresh}$) plotted against the
  strength of the burst as a fraction of total stars. The tracks are
  for models with $\log(M_*)[\msun]=10$, $10.5$, and $11$. The shaded
  regions show the range of a given model with gas fractions of
  $25$--$50\%$ just prior to the burst.}
\end{figure}

\subsection{Modeling}\label{sec:S0results2}

The lack of detectable NUV emission from all but one of the S0
galaxies shows there are not even traces of star formation in at least
97\% of SG1120 S0 galaxies. We proceed to investigate how these limits
constrain when the most recent episode of star-formation took place,
and what effect a burst of star-formation places on the constraints.

We model the S0's using the population synthesis code {\it PEGASE}
\cite[v2.0;][]{Fioc97}, and first consider a scenario in which a
galaxy has its star-formation halted completely. Our model galaxy
forms stars at a constant rate for 9.5~Gyr, roughly the age of the
Universe at $z=0.37$, and then undergoes an instantaneous burst of
star formation, after which the SFR is zero. We vary the strength of
the burst, with the models forming between 0\% to 45\% of the final
stellar mass in the burst. These burst strengths span a range from a
purely truncated disk to one that matches the median bulge-to-total
ratio found in S0's, i.e. the entire bulge forms in the burst
\citep{Christlein04}. The SFR during the pre-burst phase varies from
$\approx 1$--$15$~\msun~yr$^{-1}$, typical for galaxies at similar
stellar masses (see below) and redshift \citep[see Figure~1
of][]{Noeske07}. Within each model we set the gas fraction just prior
to the burst to be between $\approx 25\%$--50\%; the upper limit is
set by the need to convert 45\% of the gas into stars for the
strongest burst models. These SFR's and gas fractions result in
pre-burst metallicities ranging from $Z=0.5-0.8Z_\odot$. We use a
Salpeter IMF \citep{Salpeter55} and an inclination-averaged extinction
for a disk geometry. We perform this modeling with masses of
$\log(M_*)[\msun]=10$, $10.5$, and $11$, spanning the range of stellar
masses of our S0 galaxies (see Figure~\ref{optCMD}). We measure the
time ($t_{\rm thresh}$) after the burst at which the NUV emission
falls below our $2\sigma$ detection limit for the full S0 sample,
$m_{NUV}<26.0$, for $z=0.37$; $t_{\rm thresh}$ is an estimate of the
{\it minimum} time since the last significant star forming event. We
also investigate adding an additional burst (of varying strength)
earlier in the model, but its effect on $t_{\rm thresh}$ is
negligible. This is not unexpected, since the NUV emission from older
stars is well below our detection limit (see \S3.1). We show the
results of this analysis in Figure~\ref{scenarios}. As expected,
$t_{\rm thresh}$ increases with burst strength and stellar mass, with
a range over all models from $10$ to $700$~Myr.

We next investigate star-forming histories with a more gradual
reduction of star formation. Our model galaxy forms stars at a
constant rate ranging from $1$--$10$~\msun~yr$^{-1}$ and then once it
reaches $\log(M_*)=10.5$ has its SFR decline exponentially with
e-folding times ($\tau$) ranging from $0$--$2$~Gyr. We then measure
the time required for the NUV emission to fall below our detection
threshold, $t_{\rm thresh}$. From this analysis we are able to rule
out large portions of the $\tau$-$t_{\rm thresh}$ parameter space
(Figure~\ref{extsfh}); as the halting of star-formation becomes more
gradual (i.e., increasing $\tau$), the limits we place on recent
star-formation quickly exceed 1~Gyr.


\section{Discussion and Conclusion}\label{sec:discuss}

Our chief finding is a lack of NUV emission in the S0 galaxies in
SG1120, down to $m_{NUV}=26.0$, or $0.01$~\msun~yr$^{-1}$. Evidently
the S0's with masses from $\log(M_*)[\msun]\approx10$--$11$ are not
forming many new stars, but the time since their last significant star
forming episode depends on their SFH. Generally, if star formation
shut off rapidly, then they could have formed stars more recently.
Conversely, if their star formation turned off gradually, or if they
experienced a significant burst of star-formation prior to the
shut-off, then more time must have passed for them to drop below our
detection threshold. We investigate both possibilities.

\begin{figure*}
\epsscale{0.8}
\plotone{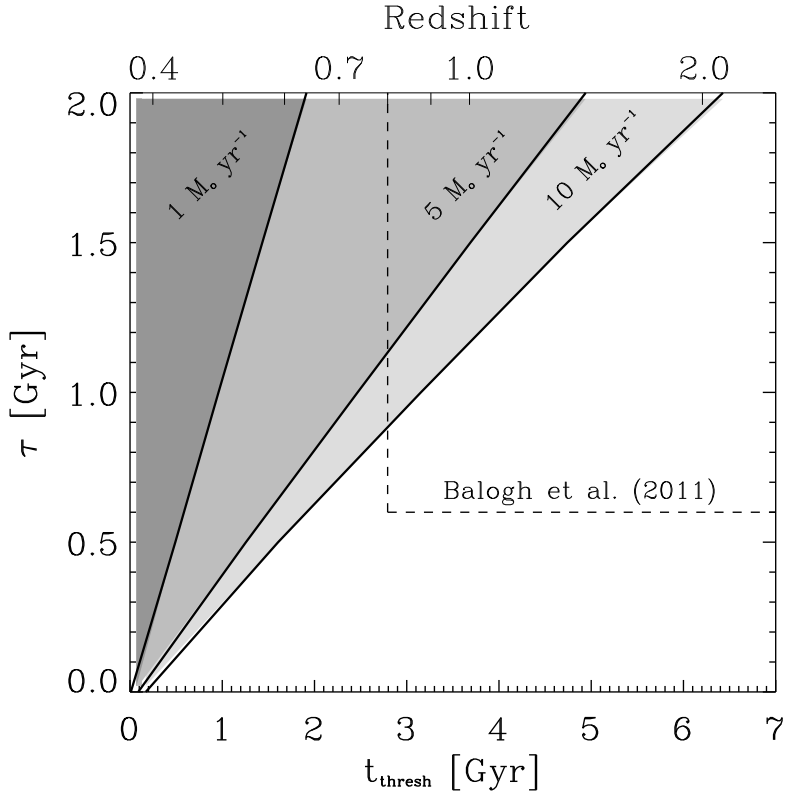}
\figurenum{5}
\caption{\label{extsfh} Plot of e-folding time ($\tau$) vs.  $t_{\rm
    thresh}$ for model galaxies with exponentially declining SFH's,
  where $t_{\rm thresh}$ is the time for the model galaxy to drop
  below our NUV detection threshold.  The model galaxies have
  $\log(M_*)[\msun]=10.5$ at the time their SFR begins to decline. The
  shaded regions, from darkest to lightest, are ruled out by our NUV
  detection limit assuming SFR's of $1$, $5$, and
  $10$~\msun~yr$^{-1}$, respectively. The dashed lines demarcate the
  parameter space considered in \cite{Balogh11}. }
\end{figure*}

In the rapid truncation scenario (Figure~\ref{scenarios}), our models
show that the minimum time since the burst ranges from $\sim0.1$ to
$0.7$~Gyr, depending on the mass of the galaxy and the strength of the
burst. While these models are consistent with the S0's having formed
at much earlier times ($>1$~Gyr), in the ``no-burst'' model the S0's
could have stopped forming stars as recently as $0.1$--$0.2$~Gyr ago,
depending on the mass. In other words, if the formation of an S0
involves a morphological transformation and a halting of star
formation, but no additional star formation, we cannot use UV
photometry to constrain meaningfully the time since that
event. However, if their formation involved an episode of significant
star formation, as one might expect in a merger, then they must have
stopped forming stars $>0.3$~Gyr prior to the time at which we are
observing them. Given the lack of E+A spectra among our S0's, which
indicate star-formation within the past $\sim1$~Gyr, it is likely that
the S0's formed at even earlier times. While the strength of the
absorption will be weaker for galaxies with no burst of
star-formation, \cite{Yang08} find E+A galaxies with burst fractions
as low as $7\%$ are consistent with their observations, demonstrating
that low burst strengths can still yield measurable E+A spectra.

We next consider the limits we can place on a gradual reduction in
SFR.  \cite{Moran07} find evidence for newly-formed S0's in groups at
the outskirts of two massive clusters at $z\sim0.5$. In the process of
forming, the SFR's of these S0's is interpreted to consist of a
gradual decline over a $\sim1$~Gyr timescale, consistent with
strangulation. In Figure~\ref{extsfh}, we model the S0's in SG1120
with similarly extended SFH's. Our S0's are consistent with a similar
slow conversion, provided that they started this decline at earlier
times. The current phase in SG1120's evolution, as the four groups
merge, is therefore unlikely to play the dominant role in S0
formation. Interestingly, a population of galaxies that lie in the
so-called ``green valley'' have been identified in groups at
$z\sim0.8$--$1$ \citep{Balogh11}, and have been interpreted as those
moving from the blue cloud to the red sequence due to an exponentially
declining SFR with $\tau\sim0.6$--$2$. These galaxies are candidate S0
progenitors given (1) their presence in groups, and (2) their
intermediate colors, since (red) S0's forming from (blue) spirals must
traverse a similar path in color space. Our models in
Figure~\ref{extsfh} have stellar masses typical of these transition
candidates. Models with an initial SFR of $1$~\msun~yr$^{-1}$ over the
full range of $\tau\sim0.6$--$2$ are consistent both with these high
redshift objects and our UV limits. Models with higher initial SFR's
begin to violate our limits for certain combinations of $\tau$ and
$t_{\rm thresh}$. Galaxies similar to these ``green valley'' group
galaxies could be the progenitors of the S0's in SG1120, but this
would again imply that the cluster assembly process is not associated
with the S0 transformation phase.

A similar picture appears to unfold at $z=0$. \cite{Hughes09} find
locally that ``green valley'' galaxies in $NUV-H$ color are
predominantly HI-deficient spirals with quenched star formation found
in higher-density environments. Further analysis has shown that these
galaxies are consistent with migration from the blue cloud to the red
sequence over at least a $\sim3$~Gyr timescale due to ram-pressure
stripping \citep{Cortese09}. A concordant result is also found over a
wider range of density \citep{Gavazzi10}.  While these results at low
redshift cannot be directly applied to higher $z$, they demonstrate
that a slow process of migration across the ``green valley'' is a
viable physical mechanism for quenching star formation, which for
SG1120 would require S0 formation prior to the cluster assembly phase.

Although the S0 fraction of SG1120 is already sufficiently large to
match that of Coma within the uncertainties and the scatter in S0
fractions, one could envision the S0 fraction of SG1120 growing by as
much as a factor of two between its current redshift and today. If so,
then S0's should be added at a rate of $\sim3$--10 per Gyr. For models
with a gradual halting in the SFR, this implies that a significant
number of S0's should be in the process of forming. However, the
likely progenitors candidates are not seen: there are $\sim 6$
non-star-forming ``passive spirals'', and at most one star-forming
S0. Hence, if strangulation is chiefly responsible for S0 formation,
then SG1120 has finished forming S0's. Conversely, if S0 formation is
ongoing in this system, then the S0's are forming without a gradual
reduction in SFR \citep[e.g.,][]{vandenBergh09}.

We find that nearly all of the S0's in SG1120 show no trace of
star-formation, and by modeling their star-formation histories with
both a rapid truncation and a gradual reduction in SFR, are able to
place limits on the time since their last significant star-forming
episode. Our constraints are weaker in the rapid reduction scenario,
particularly if S0 formation does not involve a significant burst of
star formation; from our models, the S0's could have formed stars as
recently as $\sim0.1$~Gyr ago and be consistent with our NUV limit. In
models where a burst of star-formation occurs, forming at least $20\%$
of the stellar mass, our limits imply that this occurred at least
$\sim0.3$~Gyr ago. If a more gradual reduction in star-formation
occurred, modelled as an exponentially declining SFR from a level of
$1$--$10$~\msun~yr$^{-1}$, then our limits increase to $\sim$ several
Gyr. This scenario is incompatible with SG1120 continuing to form new
S0's, as a significant number of transition galaxies would be expected
that are not observed. Evidently, the formation of S0's occurred prior
to the assembly phase of the cluster.


\acknowledgements Based on observations made with the NASA/ESA Hubble
Space Telescope, obtained [from the Data Archive] at the Space
Telescope Science Institute, which is operated by the Association of
Universities for Research in Astronomy, Inc., under NASA contract NAS
5-26555. These observations are associated with program
$\#10499$. This work is also based [in part] on observations made with
the Spitzer Space Telescope, which is operated by the Jet Propulsion
Laboratory, California Institute of Technology under a contract with
NASA. Support for this work was provided by NASA through an award
issued by JPL/Caltech GO program $\#20683$. We acknowledge financial
support for this work from NASA LTSA award NNG05GE82G and GALEX grant
NNX11AI47G. Support was also provided by NASA /HST/ G0-10499 and
JPL/Caltech SST GO-20683. K.T. acknowledges generous support from the
Swiss National Science Foundation (grant PP002-110576). D.J. would
like to thank Ann Zabludoff for helpful discussions.

{\it Facilities:} \facility{{\it GALEX}}, \facility{{\it HST} (ACS)},
\facility{{\it Magellan} (LDSS3)}, \facility{{\it Spitzer} (MIPS)},
\facility{{\it VLT} (FORS2,VIMOS)}


\begin{thebibliography}
\expandafter\ifx\csname natexlab\endcsname\relax\def\natexlab#1{#1}\fi

\bibitem[Abadi et al.(1999)]{Abadi99} Abadi, M.~G., Moore, B., \&
  Bower, R.~G.\ 1999, \mnras, 308, 947

\bibitem[Appenzeller et al.(1998)]{Appenzeller98} Appenzeller, I., et
  al.\ 1998, The Messenger, 94, 1

\bibitem[Balogh et al.(2011)]{Balogh11} Balogh, M.~L., et al.\ 2011,
  \mnras, 412, 2303

\bibitem[Bekki et al.(2002)]{Bekki02} Bekki, K., Couch, W.~J., 
\& Shioya, Y.\ 2002, \apj, 577, 651 

\bibitem[Bekki \& Couch(2011)]{Bekki11} Bekki, K., \& Couch, W.~J.\
  2011, arXiv:1105.0531

\bibitem[Bell \& de Jong(2001)]{Bell01} Bell, E.~F., \& de Jong,
  R.~S.\ 2001, \apj, 550, 212

\bibitem[Bell et al.(2003)]{Bell03} Bell, E.~F., McIntosh, D.~H.,
  Katz, N., \& Weinberg, M.~D.\ 2003, \apjs, 149, 289

\bibitem[Bertin \& Arnouts(1996)]{Bertin96} Bertin, E., \& Arnouts,
  S.\ 1996, \aaps, 117, 393

\bibitem[Blanton \& Roweis(2007)]{Blanton07} Blanton, M.~R., \&
  Roweis, S.\ 2007, \aj, 133, 734

\bibitem[Boselli et al.(2006)]{Boselli06} Boselli, A., \& Gavazzi,
  G.\ 2006, \pasp, 118, 517

\bibitem[Brown et al.(2003)]{Brown03} Brown, T.~M., Ferguson, H.~C.,
  Smith, E., Bowers, C.~W., Kimble, R.~A., Renzini, A., \& Rich,
  R.~M.\ 2003, \apjl, 584, L69

\bibitem[Byrd \& Valtonen(1990)]{Byrd90} Byrd, G., \& Valtonen, M.\
  1990, \apj, 350, 89

\bibitem[Christlein \& Zabludoff(2004)]{Christlein04} Christlein, D.,
  \& Zabludoff, A.~I.\ 2004, \apj, 616, 192

\bibitem[Cortese \& Hughes(2009)]{Cortese09} Cortese, L., \& Hughes,
  T.~M.\ 2009, \mnras, 400, 1225

\bibitem[Dale \& Helou(2002)]{Dale02} Dale, D.~A., \& Helou, G.\ 2002,
  \apj, 576, 159

\bibitem[Dale et al.(2007)]{Dale07} Dale, D.~A., et al.\ 2007, \apj,
  655, 863

\bibitem[Desai et al.(2007)]{Desai07} Desai, V., et al.\ 2007, \apj,
  660, 1151

\bibitem[Dressler(1980)]{Dressler80} Dressler, A.\ 1980, \apj, 236,
  351

\bibitem[Dressler \& Gunn(1983)]{Dressler83} Dressler, A., \& Gunn,
  J.~E.\ 1983, \apj, 270, 7

\bibitem[Dressler et al.(1997)]{Dressler97} Dressler, A., et al.\
  1997, \apj, 490, 577

\bibitem[Dressler(2004)]{Dressler04} Dressler, A.\ 2004, Clusters of
  Galaxies: Probes of Cosmological Structure and Galaxy Evolution, 206

\bibitem[Fasano et al.(2000)]{Fasano00} Fasano, G., Poggianti, B.~M.,
  Couch, W.~J., Bettoni, D., Kj{\ae}rgaard, P., \& Moles, M.\ 2000,
  \apj, 542, 673

\bibitem[Fioc \& Rocca-Volmerange(1997)]{Fioc97} Fioc, M., \&
  Rocca-Volmerange, B.\ 1997, \aap, 326, 950

\bibitem[Gavazzi et al.(2010)]{Gavazzi10} Gavazzi, G., Fumagalli, M.,
  Cucciati, O., \& Boselli, A.\ 2010, \aap, 517, A73

\bibitem[Giovanelli \& Haynes(1983)]{Giovanelli83} Giovanelli, R.,
  \& Haynes, M.~P.\ 1983, \aj, 88, 881

\bibitem[G{\'o}mez et al.(2003)]{Gomez03} G{\'o}mez, P.~L., et al.\
  2003, \apj, 584, 210

\bibitem[Gonzalez et al.(2005)]{Gonzalez05} Gonzalez, A.~H., Tran,
  K.-V.~H., Conbere, M.~N., \& Zaritsky, D.\ 2005, \apjl, 624, L73

\bibitem[Greggio \& Renzini(1999)]{Greggio99} Greggio, L., \& Renzini,
  A.\ 1999, \memsai, 70, 691

\bibitem[Gunn \& Gott(1972)]{Gunn72} Gunn, J.~E., \& Gott, J.~R., III
  1972, \apj, 176, 1

\bibitem[Han et al.(2007)]{Han07} Han, Z., Podsiadlowski, P., \&
  Lynas-Gray, A.~E.\ 2007, \mnras, 380, 1098

\bibitem[Hughes \& Cortese(2009)]{Hughes09} Hughes, T.~M., \& Cortese,
  L.\ 2009, \mnras, 396, L41

\bibitem[Icke(1985)]{Icke85} Icke, V.\ 1985, \aap, 144, 115

\bibitem[Irwin et al.(1987)]{Irwin87} Irwin, J.~A., Seaquist, E.~R.,
  Taylor, A.~R., \& Duric, N.\ 1987, \apjl, 313, L91

\bibitem[Just et al.(2010)]{Just10} Just, D.~W., Zaritsky, D., Sand,
  D.~J., Desai, V., \& Rudnick, G.\ 2010, \apj, 711, 192

\bibitem[Kautsch et al.(2008)]{Kautsch08} Kautsch, S.~J., Gonzalez,
  A.~H., Soto, C.~A., Tran, K.-V.~H., Zaritsky, D., \& Moustakas, J.\
  2008, \apjl, 688, L5

\bibitem[Kenney \& Koopmann(1999)]{Kenney99} Kenney, J.~D.~P., \&
  Koopmann, R.~A.\ 1999, \aj, 117, 181

\bibitem[Kennicutt(1998)]{Kennicutt98} Kennicutt, R.~C., Jr.\ 1998,
  \araa, 36, 189

\bibitem[Larson et al.(1980)]{Larson80} Larson, R.~B., Tinsley, 
B.~M., \& Caldwell, C.~N.\ 1980, \apj, 237, 692 

\bibitem[Lavery \& Henry(1988)]{Lavery88} Lavery, R.~J., \& Henry,
  J.~P.\ 1988, \apj, 330, 596

\bibitem[Le F{\`e}vre et al.(2003)]{LeFevre03} Le F{\`e}vre, O., et
  al.\ 2003, \procspie, 4841, 1670

\bibitem[Lewis et al.(2002)]{Lewis02} Lewis, I., et al.\ 2002, \mnras,
  334, 673

\bibitem[Martin et al.(2005)]{Martin05} Martin, D.~C., et al.\ 2005,
  \apjl, 619, L1

\bibitem[Mihos(2004)]{Mihos04} Mihos, J.~C.\ 2004, Clusters of
  Galaxies: Probes of Cosmological Structure and Galaxy Evolution, 277

\bibitem[Moore et al.(1998)]{Moore98} Moore, B., Lake, G., \& Katz,
  N.\ 1998, \apj, 495, 139

\bibitem[Moran et al.(2007)]{Moran07} Moran, S.~M., Ellis, R.~S.,
  Treu, T., Smith, G.~P., Rich, R.~M., \& Smail, I.\ 2007, \apj, 671,
  1503

\bibitem[Morrissey et al.(2005)]{Morrissey05} Morrissey, P., et al.\
  2005, \apjl, 619, L7

\bibitem[Noeske et al.(2007)]{Noeske07} Noeske, K.~G., et al.\ 2007,
  \apjl, 660, L43

\bibitem[O'Connell(1999)]{OConnell99} O'Connell, R.~W.\ 1999, \araa,
  37, 603

\bibitem[O'Donnell(1994)]{ODonnell94} O'Donnell, J.~E.\ 1994, \apj,
  422, 158

\bibitem[Papovich et al.(2003)]{Papovich03} Papovich, C., Giavalisco,
  M., Dickinson, M., Conselice, C.~J., \& Ferguson, H.~C.\ 2003, \apj,
  598, 827

\bibitem[Patel et al.(2011)]{Patel11} Patel, S.~G., Kelson, D.~D.,
  Holden, B.~P., Franx, M., \& Illingworth, G.~D.\ 2011,
  arXiv:1104.0934

\bibitem[Poggianti et al.(2009)]{Poggianti09b} Poggianti, B.~M., et
  al.\ 2009, \apj, 693, 112

\bibitem[Postman et al.(2005)]{Postman05} Postman, M., et al.\ 2005,
  \apj, 623, 721

\bibitem[Quilis et al.(2000)]{Quilis00} Quilis, V., Moore, B., \&
  Bower, R.\ 2000, Science, 288, 1617

\bibitem[Richstone(1976)]{Richstone76} Richstone, D.~O.\ 1976, \apj,
  204, 642

\bibitem[Rieke et al.(2004)]{Rieke04} Rieke, G.~H., et al.\ 2004,
  \apjs, 154, 25

\bibitem[Rix \& White(1990)]{Rix90} Rix, H.-W., \& White, S.~D.~M.\
  1990, \apj, 362, 52

\bibitem[Salpeter(1955)]{Salpeter55} Salpeter, E.~E.\ 1955, \apj, 121,
  161

\bibitem[Schlegel et al.(1998)]{Schlegel98} Schlegel, D.~J.,
  Finkbeiner, D.~P., \& Davis, M.\ 1998, \apj, 500, 525

\bibitem[Toomre \& Toomre(1972)]{Toomre72} Toomre, A., \& Toomre, J.\
  1972, \apj, 178, 623

\bibitem[Tran et al.(2005)]{Tran05} Tran, K.-V.~H., van Dokkum, P.,
  Illingworth, G.~D., Kelson, D., Gonzalez, A., \& Franx, M.\ 2005,
  \apj, 619, 134

\bibitem[Tran et al.(2009)]{Tran09} Tran, K.-V.~H., Saintonge, A.,
  Moustakas, J., Bai, L., Gonzalez, A.~H., Holden, B.~P., Zaritsky,
  D., \& Kautsch, S.~J.\ 2009, \apj, 705, 809

\bibitem[van den Bergh(1976)]{vandenBergh76} van den Bergh, S.\ 1976,
  \apj, 206, 883

\bibitem[van den Bergh(2009)]{vandenBergh09} van den Bergh, S.\ 2009,
  \apj, 702, 1502

\bibitem[van Gorkom(1996)]{vanGorkom96} van Gorkom, J.\ 1996, Cold
  Gas at High Redshift, 206, 145

\bibitem[van Gorkom(2004)]{vanGorkom04} van Gorkom, J.~H.\ 2004,
  Clusters of Galaxies: Probes of Cosmological Structure and Galaxy
  Evolution, 305

\bibitem[White et al.(2005)]{White05} White, S.~D.~M., et al.\ 2005,
  \aap, 444, 365

\bibitem[Wilman et al.(2009)]{Wilman09} Wilman, D.~J., Oemler, A.,
  Mulchaey, J.~S., McGee, S.~L., Balogh, M.~L., \& Bower, R.~G.\ 2009,
  \apj, 692, 298

\bibitem[Windhorst et al.(2002)]{Windhorst02} Windhorst, R.~A., et
  al.\ 2002, \apjs, 143, 113

\bibitem[Yang et al.(2008)]{Yang08} Yang, Y., Zabludoff, A.~I.,
  Zaritsky, D., \& Mihos, J.~C.\ 2008, \apj, 688, 945

\bibitem[Yi \& Yoon(2004)]{Yi04} Yi, S.~K., \& Yoon, S.-J.\ 2004,
  \apss, 291, 205

\bibitem[Yi et al.(2005)]{Yi05} Yi, S.~K., et al.\ 2005, \apjl, 619,
  L111

\bibitem[Zabludoff et al.(1996)]{Zabludoff96} Zabludoff, A.~T.,
  Zaritsky, D., Lin, H., Tucker, D., Hashimoto, Y., Shectman, S.~A.,
  Oemler, A., \& Kirshner, R.~P.\ 1996, \apj, 466, 104

\end{thebibliography}
\end{document}